\documentclass[11pt,preprint,nofootinbib]{revtex4}

\usepackage{amsmath,amssymb,amsfonts}
\usepackage[dvips]{graphicx}

\makeatletter

\@addtoreset{equation}{section}
\makeatother

\allowdisplaybreaks[2]

\begin{document}
\hfill KUNS-2429

\title{Spinning exact solutions with Sasakian structure \\in Gauss-Bonnet Maxwell gravity}

\author{Hiroshi Takeuchi}
\email{takeuchi@scphys.kyoto-u.ac.jp}

\affiliation{Department of Physics, Kyoto University,
Kyoto 606-8502, JAPAN}

\date{\today}

\begin{abstract}
We obtain new exact solutions in Einstein Gauss-Bonnet gravity of every odd dimension higher than three.
These new spacetimes are stationary but non-static, coupled with the Maxwell field, and asymptotic AdS at least locally.
In order to investigate such new solutions, we adopt the characteristic ansatz for the metric form.
It is presented that the local expression of our metric possesses some interesting properties, 
in which the most peculiar one is what is called Sasakian structure.
Somewhat intricate relationship is unveiled between our solution and the already-known rotating solution 
which has been only one so far in that purely gravitational theory.
We confirm the validity of the rotating spacetime with the evaluation of the finite angular momentum.
\end{abstract}

\maketitle

%\tableofcontents

\section{Introduction }

In recent decades much attention has been given to Einstein Gauss-Bonnet gravity (EGB gravity) due to modern studies of quantum gravity.  
Among the active argument about the necessity of incorporating the quantum effect to general relativity, 
the potential of EGB gravity has been indicated as the quantum correction of the low-energy effective theory originating from super string theory.
The $\alpha^{\prime}$-expansion shows that the Gauss-Bonnet term appears as the first curvature correction to Einstein gravity \cite{Gross:1986iv}. 
The past research unveiled some advantages of this theory, 
such as the ghost-free renormalizability \cite{ghostFR} or the quasi-linear property \cite{ChoquetBruhat:1988dw}.

One of the most fascinating solutions 
should describe the stationary black hole spacetime in EGB gravity, likewise in general relativity.
The static spherically symmetric black hole has been discovered in that corrected gravitational theory for a long time \cite{GBfamiliarBH}, 
studied in many stimulating investigations up to the present.
The research on this black hole solution disclosed its theoretical relationship to the one in Einstein gravity, 
beyond the similar appearance of Schwarzschild-Tangherlini solution: 
As the Gauss-Bonnet coupling goes weaker, 
one of two branches that the solution possesses gets closer to the spacetime expected from general relativity.
The extension to add the electric charge was also constructed in that gravitational theory with Maxwell electrodynamics \cite{Wiltshire:1985us}, 
resembling Reissner-Nordstr\"om solution in the previous theory.
On the other hand, studies in the black hole have revealed many peculiar features of EGB gravity absent the counterpart in Einstein gravity.
For instance the other branch of that solution goes away from general relativistic expectation in the weak coupling limit.
The massless or negative mass black holes exist on that branch without the negative cosmological constant.

Much more troublesome situation confronts us at pioneering new exact solutions.
Less exact solutions have been established than general relativity with the interference of the higher power term, 
while it has been confirmed that somewhat various solutions happen to appear \cite{Dotti:2007az, Dotti:2010bw}
only when the Gauss-Bonnet coupling was chosen to the special value \cite{Banados:1993ur}.
This trouble exposes especially at finding rotating black hole solutions.
These solutions would be gotten as the generalization of Myers-Perry solution \cite{MyersPerry} for some particular meaning, 
but such exact solutions have not been obtained yet.
Hence, investigating exact stationary solutions beyond static is important theme in EGB gravity.
In this context, it is still the unclear problem 
how far the insight accumulated in Einstein gravity is available for the present theory.
Recently, it was proven that Kerr-Schild ansatz goes wrong in EGB gravity, besides one specific exception \cite{Misao:2009kq}.
The exceptional case is just when the Gauss-Bonnet coupling is chosen specially only in five-dimensional spacetime, 
thus any extension to either higher dimensions or matter field couplings has not been found yet.
The later progress also indicates that solution possesses some defects at global behavior 
such as the absence of circularity or horizons \cite{Anabalon:2010ns, Garraffo:2008hu}.

In this paper, we obtain new exact solutions in EGB gravity of every odd dimension higher than three.
They have some good properties, such that stationary beyond static, coupled with the Maxwell field, and asymptotic AdS at least locally.
Despite these physical importance, such analytic solutions have not been known so far in EGB gravity.
We evaluate physical conserved charges of this spacetime with applying the Abbott-Deser (AD) formulation \cite{Abbott:1981ff}.
Obtaining the finite value charges, we confirm that our solution is certainly rotating.
To find out these solutions, we adopt the characteristic metric ansatz, of which the most peculiar property is called Sasakian structure \cite{Sasaki:1960}.
This mathematical object has contributed to several high-energy physics since the solution was established in Einstein gravity \cite{Gauntlett:2004yd},  
i.e. embedded in AdS/CFT \cite{Maldacena:1997re} as the Euclidean compact manifold \cite{BoyerEtAl:2008}, or 
indicated of mysterious relation to black hole physics \cite{Cvetic:2005nc,Hashimoto:2004ks}.
Our work also should develop those fields, such as applying for the compactificacion in super string theory after the suitable Wick rotation.

This paper is organized as follows.
In section 2, we provide the review of already-known solutions in EGB gravity.
In section 3, we present new exact solutions in every odd-dimension with Maxwell electrodynamics.
Sasakian structure is mentioned from the viewpoint of theoretical physics.
In section 4, physical aspects of this spacetime are revealed on five-dimension.
We clarify how to differ our solution from other solutions, and relate to them.
We also evaluate conserved charges, such as the mass, the angular momentum and the Maxwell charge.
In section 5, the paper ends with the conclusion and discussion.

\section{Special coupling choice in Einstein Gauss-Bonnet gravity }

In this section we provide the review of Einstein Gauss-Bonnet gravity 
that focused on gravitational exact solutions known previously \cite{GBfamiliarBH, Wiltshire:1985us, Misao:2009kq, Dotti:2007az, Dotti:2010bw}.
Due to the higher power term, these solutions have more complex features than ones in general relativity.
In the case that the coupling constant of that term is specially chosen, somewhat various solutions have been constructed so far.
In particular, the only one rotating solution found previously in five-dimension also needs this condition \cite{Misao:2009kq}.
We explain a specific feature of that rotating solution and its non-trivial relation to static topological black hole solutions.

In the D-dimensional spacetime, EGB gravity with Maxwell electrodynamics is described by the following action,
\begin{align}
 S =& \int \sqrt{|g|}\text{d}x^{D} \Bigl[\tfrac{1}{16\pi G}\bigl( R +\alpha \mathcal{L}_{GB}- 2 \Lambda \bigr)
	- \tfrac{1}{4\pi g^2_c}F_{\mu \nu}F^{\mu \nu} \Bigr], \label{action}\\
	=& S_{\text{EGB}} + S_{\text{Maxwell}}, \\
 & \mathcal{L}_{GB} = R^2 - 4 R_{\mu \nu}R^{\mu \nu} + R_{\mu \nu \rho \sigma}R^{\mu \nu \rho \sigma},
\end{align}
with $\alpha $ the coupling constant of the Gauss-Bonnet term, $\Lambda $ the cosmological constant, and $F = \text{d}A$ the field strength of Maxwell field.
The gauge or gravitational coupling constant is respectively written as $g_c$ or $G$.
Equations of motion (EOMs) read
\begin{align}
 G_{\mu \nu } + \alpha H_{\mu \nu } + \Lambda g_{\mu \nu } & = (16\pi G)~ T_{\mu \nu}, \label{GBEOM}\\
 \nabla  _{\mu} F^{\mu \nu} & = 0,
\end{align}
where
\begin{align}
 G_{\mu \nu } &=R_{\mu \nu } - \frac{1}{2}g_{\mu \nu }R,\\
 H_{\mu \nu } &=2(R R_{\mu \nu }-2 R_{\mu \rho }R^{\rho}_{~ \nu}-2 R^{\rho \sigma}R_{\mu \rho \nu \sigma}
+R_{\mu \rho \sigma \lambda }R_{\nu}^{~ \rho \sigma \lambda })-\frac{1}{2}g_{\mu \nu }\mathcal{L}_{GB} ,\\
 T_{\mu \nu } &= \frac{1}{2 \pi g^2_c}\bigl( F_{\mu \lambda }F_{\nu }^{~\lambda }-\frac{1}{4}g_{\mu \nu}F_{\lambda \rho }F^{\lambda \rho}\bigr).
\end{align}
Hereafter, we set the gauge coupling constant to $g_c=4 \sqrt{G}$ without loss of generality, 
absorbing it into the scale of the matter flux $F$.

In this theory, there is the familiar black hole solution which can be seen as the counterpart of 
Reissner-Nordstr\"om solution in general relativity \cite{GBfamiliarBH, Wiltshire:1985us}.
The solution is written in the form,
\begin{align}
\text{d}s^2
= -R(r)&\text{d}t^2 + \frac{\text{d}r^2}{R(r)}
+r^2 \text{d} \Omega   _{D\text{-}2}^2, \label{familiarBH}\\
 R(r) &= 1+\frac{r^2}{2\tilde{\alpha }}\Bigl\{ 1 \mp \sqrt{1 + 4\tilde{\alpha }\Bigl( \tfrac{M }{r^{D-1}}
+\tfrac{2 \Lambda }{(D-1)(D-2)}-\tfrac{Q^2}{8(D-3)(D-2)r^{2D-4}} \Bigr)} ~ \Bigr\}, \label{familiarBHfunc}\\
 A_{\mu} \text{d}x^{\mu} &= \frac{Q}{r^{D-3}}\text{d}t,
\end{align}
with $\text{d} \Omega   _{D\text{-}2}^2$ the unit (D-2)-dimensional sphere metric, $\tilde{\alpha }= (D-3)(D-4)\alpha$, 
$M$ the mass parameter, and $Q$ the electric charge parameter.
This spacetime is static, spherically symmetric, and asymptotically (A)dS or flat.
The solution has two branches: The minus sign case is called the Einstein branch 
because it is connected to the solution of general relativity in the weak coupling region.

As is the massless and chargeless case of (\ref{familiarBH}), EGB gravity allows two maximally symmetric vacua generically.
Note that if these vacua degenerate, something special occurs in the gravitational solution space.
This occurs when you choose the coupling constant as the following special value
\begin{align}
 \alpha  = - \frac{(D-1)(D-2)}{8(D-3)(D-4)\Lambda } \label{SpCa}.
\end{align}
It leads the metric function $R(r)$ with vanishing electric charge into a fairly simpler form
\begin{align}
R(r) = \frac{r^2}{l^2} +1 \mp \bigl( \frac{2 M}{l^2 r^{D-5}} \Bigr)^{\tfrac{1}{2}} ,~~~~ l^2 = - \frac{(D-1)(D-2)}{4 \Lambda} \label{SpCR}.
\end{align}
The special choice probably induces the enhancement on the gravitational solution space,
thus some noteworthy solutions happen to appear only at this choice of parameters \cite{Misao:2009kq, Dotti:2007az, Dotti:2010bw}.
Particularly, in five-dimension more assorted solutions have been found than in the other dimension.
The five-dimensional curiousness of the choice can be read from either the black hole metric function (\ref{SpCR}) 
whose mass parameter can absorb the contribution of the internal (D-2)-dimensional space, 
or the action getting the peculiar name called the Chern-Simon case in Lovelock gravity \cite{Banados:1993ur}.

One of such five-dimensional solutions has been discovered by using the Kerr-Schild ansatz \cite{Misao:2009kq}. 
The solution has two rotational parameters and no electric charge with any matters.
It is known that the solution has a specific feature:
When these rotational parameters coincide, the line element can be taken to the non-trivial static chart.
In the Kerr-Schild form, the line element \cite{Misao:2009kq} whose rotational parameters are adjusted to equal is given by
\begin{align}
 \text{d}s^2
 = f(r)&(h_{\mu}\text{d}x^{\mu})^2+\text{d}\bar{s}^2, \label{misaoKS}\\
 h_{\mu}\text{d}x^{\mu}
&= \frac{1}{\Xi }(\text{d}t +\frac{a}{2}(\text{d}\psi-\cos \theta \text{d}\phi)) +\frac{r^2 \text{d}r}{(\tfrac{r^2}{\mathcal{L}^2}+1)(r^2+a^2)},\\
 f(r) &= (\tfrac{1}{\mathcal{L}^2}-\tfrac{1}{l^2})(r^2+a^2),
\end{align}
where $\text{d}\bar{s}^2$ is the seed line element of the AdS spacetime,
\begin{align}
 \text{d}\bar{s}^2
 = & -\frac{(\tfrac{r^2}{\mathcal{L}^2}+1)}{\Xi}\text{d}t^2+\frac{r^2 \text{d}r^2}{(\tfrac{r^2}{\mathcal{L}^2}+1)(r^2+a^2)}
+\frac{(r^2+a^2)}{4\Xi}(\text{d}\theta ^{2} +\sin ^2 \theta  \text{d}\phi^{2} +(\text{d}\psi - \cos \theta \text{d}\phi) ^2),\\
 & \Xi = 1-\frac{a^2}{\mathcal{L}^2},
\end{align}
$\mathcal{L}$ is the seed AdS radius independent of the special coupling parameter $l$, 
and $a$ is the oblateness parameter into which two rotational parameters degenerate.
It can be shown that the line element (\ref{misaoKS}) is diffeomorphic to the following one
\begin{align}
\text{d}s^2
 = & -(\tfrac{\tilde{r}^2}{l^2}-\tfrac{a^2}{l^2}+1)\text{d}\tilde{t}^2 + \frac{\text{d}\tilde{r}^2}{(\tfrac{\tilde{r}^2}{l^2}-\tfrac{a^2}{l^2}+1)}
+\frac{\tilde{r}^2}{4\Xi}(\text{d}\theta ^{2} +\sin ^2 \theta  \text{d}\phi^{2} +\frac{\Xi_l}{\Xi}(\text{d}\tilde{\psi}-\cos \theta \text{d}\phi) ^2), \label{misaoSC}\\
 & \Xi _l = 1-\frac{a^2}{l^2}.
\end{align}
On this static chart, the three-dimensional space with $\tilde{t}$ and $\tilde{r}$ fixed is not the round sphere 
but the squashed one over the $U(1)$ direction, i.e. the squashed $S^3$.
Therefore, the adjusted solution (\ref{misaoKS}) is not identical to the familiar static black hole (\ref{familiarBH}) 
but a sort of so-called topological black holes \cite{Cai:2001dz}.
Seeing it as the static black hole, parameters $a$ and $\mathcal{L}$ correspond to the mass and squashing rate after some recombination.
Its asymptotic AdS radius is described by $l$ rather than $\mathcal{L}$.

In the vacuum case of EGB gravity, static topological black holes were constructed in arbitrary coupling values and general spacetime dimensions \cite{Dotti:2007az, Dotti:2010bw}.
The solution is given by replacing the internal sphere $\text{d} \Omega   _{D\text{-}2}^2$ of (\ref{familiarBH}) to the other (D-2)-dimensional space $\text{d} s  _{D\text{-}2}^2$
selected by EOM constraints, thus most of them possess some artificial boundary structure.
Usually these constraints are quite harder than in Einstein gravity, restricting not only the Ricci tensor but also the Weyl tensor.
However, it has been found that in the case of the special choice, static topological black hole solutions extend to much larger class 
because the choice considerably relaxes the EOM requirement \cite{Dotti:2007az}.
In particular, the relaxation in five-dimension is so prominent that any Euclidean line element $\text{d} s _{3}^2$ is permitted for the internal three-dimensional space. 
Therefore, the solution (\ref{misaoKS}) belongs to two families of gravitational solutions, 
one of which is the two-rotating spacetime, and the other is the enhancement of static topological black holes.
It does not seem so unnatural that the solution has more families other than them, which has been unknown so far.

Whereas the higher dimensional static topological black holes were researched \cite{Dotti:2010bw, Oliva:2012ff}, 
each of these two families still has the problem how far it can generalize to more other directions.
The research for topological black holes shows that the prominent relaxation to any internal (D-2)-dimensional space happens at the Chern-Simon case of Lovelock gravity.
On the other hand, the higher dimensional rotating solution has not been found at all.
Moreover, both of them have not been deformed with coupling any matters such as Maxwell field in the exact manner.

\section{Charged spinning exact solution in odd-dimension}

In this section, new exact solutions in every odd dimension higher than three are investigated 
in EGB gravity with Maxwell electrodynamics.
We present our ansatz to find new solutions, and reveal its good properties.  
Explaining some technical goodness, we derive new solutions.
The most novel property is Sasakian structure, which is a mathematical object but has some useful aspects in theoretical physics.
Since we interpret it from the latter point of view, the description slightly lacks some mathematical exactness.
After that, we clarify its stationarity and non-staticity except for the specific case.

In order to investigate new solutions in Gauss-Bonnet Maxwell theory on arbitrary (2n+1)-dimensional spacetime, 
we adopt the ansatz restricting the metric to
\begin{align}
 \text{d}s^2
=& -(\text{d}\tau + \lambda  y \mathcal{A})^2+ \frac{\text{d}y^2}{Y(y)}+Y(y)\mathcal{A}^2+y\ \text{d}\tilde {s}^2 , \label{sasaAZ}\\
  & \mathcal{A} = \text{d}\varphi + \tilde {A},
\end{align}
with $Y(y)$ an arbitrary function of $y$, $\lambda $ a nonzero constant parameter, 
and $\text{d}\tilde {s}^2$ the Fubini-Study metric of the internal $CP_{n \text{-}1}$ space\footnotemark. 
\footnotetext{As is the case of topological black holes, there are some variations of the internal space besides $CP_{n \text{-}1}$.
A part of them is expressed by $\gamma$: 
$\gamma =0$ is flat space, and $\gamma^2 < 0$ is the negative curvature homogeneous space analogous to $CP_{n \text{-}1}$.}
We normalize this 2(n-1)-dimensional space as
\begin{align}
\mathrm{Ricci}_{C P_{n \text{-}1}}= \frac{n \gamma ^2}{2} \text{d}\tilde {s}^2, ~~~~\text{d}\tilde {A}= \tilde {J},
\end{align}
where $\tilde {A}$ is the vector potential of K\"ahler form $\tilde {J}$ on the internal 2(n-1)-dimensional space $\text{d}\tilde {s}^2$.

This class of metrics (\ref{sasaAZ}) has some proficient properties.
Let us show them in order.
First of all, we present some technical goodness.
On this ansatz, there is an assured Maxwell solution.
The following vector potential satisfies the Maxwell equation regardless of the concrete form of $Y(y)$, 
\begin{align}
  A_{\mu} \text{d}x^{\mu} =& \frac{Q}{y^{n-1}}\mathcal{A}. \label{MxAZ}
\end{align}
For this reason, switching on the matter flux, we assign the electromagnetic charge as this.

To solve the gravitational equation of motion (\ref{GBEOM}), the technical chart (\ref{sasaAZ}) exerts its power.
Using the following vielbein $\{e^A_{\mu}\mid A=(0, a, 2n\text{-}1, 2n) \}$, 
\begin{align}
& e^{0} = \text{d}\tau + \lambda y \mathcal{A},~~~~
  e^{a} = \sqrt{y}~ \tilde{e}^{a}~,~~~~
 \tilde{e}_{p}^{a} \tilde{e}_{q}^{a} \text{d}x^p \text{d}x^q = \text{d}\tilde {s}^2, \notag \\
& e^{2n\text{-}\! 1} = \frac{\text{d}y}{\sqrt{Y(y)}},~~~~
 e^{2n} = \sqrt{Y(y)}~ \mathcal{A},
\end{align}
with $a = 1,...,2n \text{-} 2$ and $\{x^p \}$ the coordinate of $CP_{n \text{-}1}$,
the Riemann tensor gets a considerably simple expression,
\begin{align}
&R^{0C}_{~~~AB } =-\frac{\lambda^2}{2} \delta^{0}_{[A} \delta^{C}_{B ]},~~~~~~
 R^{2n\text{-}\! 1 2n}_{~~~~~~AB } = \bigl( \frac{3}{2}\lambda ^2 -Y^{\prime \prime}(y) \bigr) \delta^{2n\text{-}\! 1}_{[A} \delta^{2n}_{B ]} 
 +\frac{1}{2} \bigl( \lambda ^2 +\mathcal{N}_2 \bigr) \tilde{J}_{ab} \delta^{a}_{[A} \delta^{\text{b}}_{B ]},\\
& R^{c2n\text{-}\! 1}_{~~~AB } = \frac{\mathcal{N}_2}{2} \delta^{c}_{[A} \delta^{2n\text{-}\! 1}_{B ]} +\frac{1}{2} \bigl( \lambda ^2 +\mathcal{N}_2 \bigr) \tilde{J}_{~b}^c \delta^{b}_{[A} \delta^{2n}_{B ]} ,\\
& R^{c2n}_{~~~AB } = \frac{\mathcal{N}_2}{2} \delta^{c}_{[A} \delta^{2n}_{B ]} -\frac{1}{2} \bigl( \lambda ^2 +\mathcal{N}_2 \bigr) \tilde{J}_{~b}^c \delta^{b}_{[A} \delta^{2n\text{-}\! 1}_{B ]} ,\\
& R^{cd}_{~~AB } = \frac{\mathcal{N}_1}{2} \delta^{c}_{[A} \delta^{d}_{B ]} +\frac{1}{2} \bigl( \lambda ^2 +\mathcal{N}_1 \bigr) \bigl( \tilde{J}^c _{~[a}\tilde{J}^d _{~b]} +\tilde{J}^{cd}\tilde{J}_{ab} \bigr) \delta^{a}_{[A} \delta^{b}_{B]}
+\bigl( \lambda ^2 +\mathcal{N}_2 \bigr) \tilde{J}^{cd} \delta^{2n\text{-}\! 1}_{[A} \delta^{2n}_{B ]},\\
&~~ \mathcal{N}_1 = \frac{\gamma ^2}{y} -\frac{1}{y^2}Y(y),
~~~~ \mathcal{N}_2 = \frac{1}{y^2} Y(y) -\frac{1}{y}Y^{\prime}(y),
\end{align}
As above, the 0th direction of the Riemann tensor is coincident with the maximal symmetric case, 
and the $(2n\text{-}1)$ or $2n$th direction is interchangeable up to the sign.
Every component is described in systematical forms, only by $\delta $ and $\tilde{J} $.
These simplicity results in diagonalizing EOMs (\ref{GBEOM}) as well as the Ricci tensor.
In addition the diagonalization accompanies specific degeneracies, 
that is, at most three different eigenvalues are allowed respectively on the $(0,0)$, $(2n\text{-}1,2n\text{-}1)\text{=}(2n,2n)$ and $(a,a)$ component.
All these components are ordinary differential equations of $y$, mutually correlated by the conservation law.
In the case of special coupling choice, these equations are further reduced to two equations with the choice $\lambda=\tfrac{2}{l}$.
The one equation comes from its trace part
\begin{align}
& G^A_{~A} + \frac{l^2}{4(n-1)(2n-3)} H^A_{~A} - \frac{n(2n-1)(2n+1)}{2l^2} - (16\pi G)~ T^A_{~A} ,  \notag \\
=& n\bigl( 1 +\frac{l^2}{4y^2}Y(y)-\frac{l^2}{4y}Y^{\prime}(y)\bigr) \bigl( -\frac{4 (2n-3)}{l^2} -\frac{2(n-2)\gamma ^2}{y} +\frac{(2n-5)}{y^2}Y(y) +\frac{1}{y}Y^{\prime}(y) \bigr) \notag \\
& + n\bigl( \frac{\gamma ^2 l^2}{4y}+1-\frac{l^2}{4y^2}Y(y)\bigr) \bigl( (n-1)(\tfrac{n}{2}-1)(-\frac{\gamma ^2}{y} -\frac{4}{l^2} +\frac{1}{y^2}Y(y)) -\frac{8}{l^2} +Y^{\prime \prime}(y)\bigr) \notag \\
& + \frac{n(n-1)(2n-3) Q^2}{4 y^{2n}}, \notag \\
= & ~0, \label{traceEOM}
\end{align}
and the other from the $(2n\text{-}1, 2n\text{-}1)$ component of EOMs,
\begin{align}
& G_{2n\text{-}\! 1 2n\text{-}\! 1} + \frac{l^2}{4(n-1)(2n-3)} H_{2n\text{-}\! 1 2n\text{-}\! 1} - \frac{n(2n-1)}{2l^2} - (16\pi G)~ T_{2n\text{-}\! 1 2n\text{-}\! 1} \notag \\
=& \frac{n(n-2)}{(2n-3)}\bigl( \frac{\gamma ^2 l^2}{4y}+1-\frac{l^2}{4y^2}Y(y)\bigr) \bigl( \frac{(1-n)\gamma ^2}{2y} -\frac{2(n+1)}{l^2 } +\frac{(n-3)}{2y^2}Y(y)+\frac{1}{y}Y^{\prime}(y) \bigr) \notag \\
&- \frac{(n-1)(n-2) Q^2}{4 y^{2n}}, \notag \\
= & ~0. \label{conEOM}
\end{align}
All the other components can be written down in linear combinations of (\ref{traceEOM}) and (\ref{conEOM}).
Here, we find that the following metric function satisfies the trace equation (\ref{traceEOM}), 
\begin{align}
Y(y) = \frac{4}{l^2} y^2 + \gamma ^2 y \pm \sqrt{\mu_1~ y^{3-n}+\mu_2~ y^{4-n} +\frac{2(2n-3)}{n l^2}Q^2~ y^{4-2n}} \label{SOL},
\end{align}
with $\mu_1$, $\mu_2$ constant parameters and $l$ the special coupling parameter as (\ref{SpCR}).
In the case of $n=2$, this is the new solution itself 
because the residual equation (\ref{conEOM}) automatically vanishes by the overall factor. 
Note that it is just the exposure of the five-dimensional curiousness.
In $n\geq 3$ the residual constraint, being the rank-one ordinary differential equation, inhibits one integration constant.
As a result, the higher dimensional solution is (\ref{SOL}) with $\mu_2=0$.

The most peculiar property we explain next is what is called Sasakian structure \cite{Sasaki:1960}
(See the reference \cite{BoyerEtAl:2008}, about recent studies).
This is interpreted in a bit mathematical context, while it could be seen as symbolizing the advent of the technical goodness.
Roughly speaking, Sasakian structure is characterized by the specific Killing vector field $\zeta $ that must have a constant norm.
Besides these conditions, $\zeta $ needs to fulfill the following condition\footnotemark
\begin{align}
 C_{r} R_{\mu \nu \rho \lambda }\zeta^{\lambda} = g_{\rho \mu }\zeta _{\nu } - g_{\rho \nu }\zeta _{\mu } , \label{sasaCON}
\end{align}
with a certain constant $C_{r}$.
\footnotetext{In the literature $C_{r}$ is often fixed a certain value, but we do not so for convenience.
The signature of $C_{r}$ depends on whether $\zeta$ is space-like or time-like.}
This characteristic vector field $\zeta $ is often called the Reeb vector field.
Generically, the metric with Sasakian structure can be seen as a $U(1)$-bundle 
whose fiber direction corresponds to the one-form $\zeta ^{\flat}$ dual of the Reeb vector $\zeta $.
The one-form $\zeta ^{\flat} $ induces the K\"ahler structure on the 2n-dimensional base space at least locally.
Restricting to be on the base space, its exterior derivative $\text{d}\zeta ^{\flat}$ can be regarded as the K\"ahler form of the 2n-dimensional space.

In our ansatz (\ref{sasaAZ}), the vector field $\partial _{\tau}$ can be identified as the Reeb vector $\zeta $.
It satisfies (\ref{sasaCON}) as well as all the other necessary conditions.
Some known Einstein spacetimes are recovered with tuning the metric function.
The following choice satisfies the gravitational EOM (\ref{GBEOM}) 
of the action (\ref{action}) without the Gauss-Bonnet term and the Maxwell term, 
\begin{align}
 Y(y) = \lambda ^2 y^2 + \gamma ^2 y + \frac{M}{y^{n-1}},~~~~\lambda = \sqrt{\frac{-2\Lambda}{n(2n-1)}}, \label{EinY}
\end{align}
The case of $M=0$ corresponds locally to the AdS spacetime.
In five-dimension this solution has been gotten by a certain supersymmetric limit \cite{Cvetic:2005nc}.
Its Euclidean counterpart \cite{Gauntlett:2004yd} was formerly constructed.
This manifold is called $Y^{pq}$, linked to (\ref{EinY}) locally through the suitable Wick rotation $(\tau \rightarrow \text{i} \tau,~ \lambda \rightarrow \text{i} \lambda )$.
The investigation around this Euclidean Sasaki-Einstein manifold has been developed a lot in studies of high energy physics.
For example, it has been generalized to both higher dimensions \cite{Gauntlett:2004hh} and lower isometry \cite{Cvetic:2005ft}.
However, few exact solutions with Sasakian structure have been found so far in some physical theory beyond purely Einstein gravity.
Actually, even in that theory with Maxwell electrodynamics the above ansatz (\ref{sasaAZ}), (\ref{MxAZ}) cannot catch the solution, 
then no exact spinning solution with matter flux has been constructed around that condition (\ref{sasaAZ}).
The difficulty arises from taking all the back reactions into account, 
so that it may need to deform the action itself as well as the metric ansatz.
Recently, in five-dimensional gauged minimal supergravity theory, which can be seen as adding the Chern-Simon term to Einstein Maxwell theory, 
the exact solution of the deformed metric ansatz from (\ref{sasaAZ}) has been discovered \cite{Houri:2012zv}.
It is interesting that the similar problem has occurred at the generalization of the Kerr-Newman black hole: 
The higher dimensional rotating black hole in Einstein Maxwell theory has not been discovered as the exact solution yet, 
whereas five-dimensional rotating black hole solution in gauged minimal supergravity has been found exactly \cite{Chong:2005hr}.
It is the delicate problem whether a rotating solution can be coupled with the Maxwell field in the exact form.

Here we explain how the non-static property goes on in our metric form:
The static condition cannot be fulfilled by any tune of the time-like Killing vector field, 
besides the specific exception of the metric function $Y(y)$. 
Whether the spacetime is static or not is due to the existence of the time-like Killing vector field $\xi$ 
that satisfies the following condition everywhere in that spacetime, 
\begin{align}
\xi^{\flat} \wedge \text{d}\xi^{\flat} =0,
\end{align}
According to our metric ansatz (\ref{sasaAZ}), 
Killing vector fields are included in $SU(n)\times U(1)\times \text{{\boldmath $R$}}$.
Considering the property of the Reeb vector, the everywhere time-like Killing direction, any $\xi$ can be written in
\begin{align}
\xi = & \partial _{\tau}+\Omega \partial _{\varphi } +\xi_{SU(n)},
\end{align}
where $\Omega $ is an arbitrary constant, $\xi_{SU(n)}$ is any combination of $SU(n)$ Killing vector fields originated from the internal $CP_{n \text{-}1}$.
The exterior derivative of its dual one-form reads
\begin{align}
\text{d}\xi^{\flat}= & \bigl\{ \Omega (\tfrac{1}{y}Y(y)-\lambda^2 y)-\lambda \bigr\} J +\Omega~ \text{d}(\tfrac{1}{y}Y(y)-\lambda^2 y)
\wedge (y\mathcal{A}) +\text{d}\xi^{\flat}_{SU(n)} \label{static-eq},
\end{align}
with $J$ the K\"ahler form on 2n-dimensional base space, $J = \text{d}(y \mathcal{A})$.
At this point, the exceptional static case can be read from this formula.
When the metric function fixes $Y(y)=\lambda^2 y^2 + C_v y$ with any constant $C_v$, the second term vanishes automatically. 
The first term is also able to cancel out with choosing $\Omega =\frac{\lambda }{C_v} $.
Hence the static condition is satisfied without $\xi_{SU(n)} $.
In any other cases, however, first two terms of (\ref{static-eq}) must remain on some components of $\xi^{\flat} \wedge \text{d}\xi^{\flat}$.
This contains the term $\mathcal{A}\wedge J$, which is the $SU(n)$ isometry singlet. 
Therefore even if you turn on $\xi_{SU(n)}$, the vector representation of $SU(n)$, it does not cancel that term but raises rather some terrible components.

The exceptional static case above can be checked by the concrete form.
After changing the coordinate as follows,
\begin{align}
\tau = & \sqrt{\frac{\lambda }{\Omega }}~t,~~~~\varphi=\phi +\sqrt{\lambda \Omega }t, \label{BLtrf}
\end{align}
the metric with $Y(y)=Y_{\lambda}(y) + C_v y$ is given by 
\begin{align}
\text{d}s^2
 = & - (\lambda^2 y + C_v)\text{d}t^2 + \frac{\text{d}y^2}{\lambda^2 y^2 + C_v y} 
+ y~ (\text{d}\tilde {s}^2 + \frac{\lambda }{\Omega }(\text{d}\phi + \tilde {A})^2),
\end{align}
If the similar transformation is done in any other metric function, the line element seems like Boyer-Lindquist coordinate \cite{BoyerLindquist}.
It was analyzed that the existence of  Boyer-Lindquist coordinate relates the circularity of the metric, 
which our solution possesses but not the previous rotating solution \cite{Anabalon:2010ns}.
It is remarkable that the non-trivial static case above cannot occur in the present solution (\ref{SOL}) unless the parameters are tuned in five-dimension.

\section{5-dimensional solution and conserved charges}

In this section, we show some aspects of our solutions on five-dimensional Gauss-Bonnet Maxwell theory.
Because in five-dimension, several solutions have been discovered as in section 2, 
we explain how to differ our solution from other solutions, and relate ours to others.
The global structure of the new spacetime is also argued briefly.
We evaluate conserved charges with applying the AD formulation \cite{Abbott:1981ff}.
The mass, the angular momentum and the Maxwell charge are obtained as well-defined values.

In five-dimensional spacetime, the present exact solution can be written in the following form with fixing $\gamma = 2$,
\begin{align}
\text{d}s^2 = & -(\tfrac{1}{l^2} r^2 + 1 )\text{d}t^2 + \frac{V(r)}{16}(\tfrac{2 }{l }\text{d}t + \text{d}\phi + \cos{\theta} \text{d}\psi)^2 
+ \frac{\text{d}r^2}{\frac{1}{l^2} r^2 + 1+\tfrac{V(r)}{4 r^2}} \notag \\
&~~~~ + \frac{r^2}{4} (\text{d} \theta ^2 +\sin{\theta}^2\text{d}\psi ^2 +(\text{d}\phi +\cos{\theta} \text{d}\psi)^2) ,\label{5Dsol}\\
 V(r)& = \pm \sqrt{\mu_1 r^2+\mu_2 r^4 +\tfrac{Q^2}{l^2}},\\
 A_{\mu} \text{d}x^{\mu}& = \frac{Q}{4r^2}(\tfrac{2 }{l }\text{d}t + \text{d}\phi + \cos{\theta} \text{d}\psi),
\end{align}
In this expression the isometry $SU(2) \times U(1) \times \text{{\boldmath $R$}}$ is constituted from 
the internal $S^2$ space, the axial symmetry $\partial _{\phi}$ and the time translation $\partial _t$.
This coordinate connects the previous one (\ref{sasaAZ}), (\ref{MxAZ}) through the transformation of (\ref{BLtrf}) with $\Omega=\frac{1}{2l } $ 
and $(y\rightarrow r^2,~ t \rightarrow \frac{t}{2},~ \phi \rightarrow \frac{\phi}{4} )$.
Composed of  the $S^2$ and $U(1)$, the round $S^3$ part is exposed at the last term of (\ref{5Dsol}).
Chosen to vanishing $Q$ and $\mu_1$, 
the solution turns to get the static chart with the other $\Omega$ choice.
In this case, the solution comes in a part of the squashed $S^3$ black hole solution (\ref{misaoSC}), 
in which the mass and squashing parameter are decided by one parameter $\mu_2$.
However, in all the other cases, our solution cannot become static, then they differ from static topological black holes.
It is notable that the squashing parameter $\mu_2$ of five-dimension is prohibited in higher dimensional generalization.

The five-dimensional solution is also different from the previous rotating solution \cite{Misao:2009kq} even if the electromagnetic charge $Q$ vanishes, except for the static case.
The previous solution has the above large isometry only when its two rotational parameters coincide, 
as is the case of Myers-Perry solution or its generalization with the cosmological constant \cite{MyersPerry, Gibbons:2004uw} in general relativity.
In Kerr-Schild form, their solution consists of the same null geodesic vector as five-dimensional Kerr-AdS solution \cite{Misao:2009kq, Gibbons:2004uw}.
Thus, once the coincidence breaks into two angular momentum, they decrease the isometry to $U(1)\times U(1)\times \text{{\boldmath $R$}} $.
Moreover, there is no Killing vector field possessing the constant norm in the previous solution besides the static case.
This means the absence of Sasakian structure on their solution, then it concludes the difference from our solution.
What this variation around non-static solutions means physically in EGB gravity is not so obvious, but much interesting subject.
EGB gravity might allow somewhat larger class of rotating solutions than expected from the knowledge accumulated in general relativity.

Considering the global structure, you can see the coordinate $r$ almost corresponds to the radial direction.
As $r \rightarrow +\infty$, the Riemann tensor of our solution gets closer to the one on the AdS spacetime, 
that is, $R^{AB }_{~~~CD }\rightarrow - \frac{2}{l^2} \delta^{A } _{[C }\delta^{B } _{D ]}$.
For this reason, our spacetime is regarded as asymptotically AdS at least locally \cite{Misao:2009kq}.
Tuning parameters could raise the Killing horizon on the positive finite radius of that spacetime.
The Killing vector field $\partial _t$ has the norm $\bigr( \tfrac{1}{4l^2}V(r)-\tfrac{1}{l^2} r^2 - 1 \bigl) $, which allows some nodes.
Even when the Maxwell charge $Q$ vanishes, residual two-parameters can cause the extremal case.
Furthermore, this solution has the curvature singularity at $r = 0$.
Some curvature invariants diverge there, for instance the Kretschmann invariant $K=R^{\mu \nu \rho \sigma }R_{\mu \nu \rho \sigma }$ ;
\begin{align}
K=\frac{40}{l^4}+\frac{8V(r)^2}{r^8} -\frac{4V(r)V^{\prime}(r)}{r^7} +\frac{3V^{\prime}(r)}{l^2 r^3}+\frac{17 V^{\prime}(r)^2}{16r^6}
 +\frac{V^{\prime \prime}(r)}{l^2 r^2} - \frac{V^{\prime}(r)V^{\prime \prime}(r)}{8r^5}+\frac{V^{\prime \prime}(r)^2}{16r^4}.
\end{align}
There is one more possibility of singularity at the node of $V(r)$, attributed to its derivative term with the square root form.
However, the similar situation occurs in the non-Einstein branch of the static black hole solution (\ref{familiarBHfunc}), 
avoided in the Einstein branch owing to some physical condition.
Thus it is expected that our solution also does on the precisely global analysis.
The present coordinate might be improved after the global inspection  more cautiously and the adjustment from some physical requests.
All these structures except for $\mu_2$ are preserved to the higher dimensional generalization.

Finally let us evaluate the physical conserved values of this spacetime around the AdS background.
In order to face this higher curvature gravity theory, we adopt the AD formulation 
which has been applied to EGB gravity by S. Deser {\it et al.} \cite{Abbott:1981ff}.
Using the background metric $\bar{g}_{\mu \nu}$, the metric is decomposed into
\begin{align}
& g_{\mu \nu} = \bar{g}_{\mu \nu} + h_{\mu \nu},
\end{align}
where $\bar{g}_{\mu \nu}$ has the curvature $\bar{R}^{\mu \rho}_{~~\nu \sigma } = -\frac{2}{l^2} \delta^{\mu }_{[\nu } \delta^{\rho }_{\sigma ]}$,
and the deviation $h_{\mu \nu}$ has the information of the present solution.
For extracting the spacetime energy-momentum tensor $T^{(h)}_{\mu \nu}$, it needs to expand the gravitational field equation by $h$.
When the coupling constants are tuned specially with $l$, EOM (\ref{GBEOM}) is factorized into the remarkably simple form
\begin{align}
 (16\pi G) \frac{\delta S_{\text{EGB}} }{\delta g^{\mu \nu}} & =
 15 l^2 g_{\mu \nu_5}\delta ^{[ \nu_1}_{[ \nu} \check{R}^{\nu_2 \nu_3}_{~~~~\nu_1 \nu_2 } \check{R}^{\nu_4 \nu_5 ]}_{~~~~\nu_3 \nu_4 ]}
 = (16\pi G)~ T_{\mu \nu}, \label{spC_EOM}\\
 \check{R}^{\mu \rho}_{~~\nu \sigma } &= R^{\mu \rho}_{~~\nu \sigma } +\frac{2}{l^2} \delta^{\mu }_{[\nu } \delta^{\rho }_{\sigma ]}.
\end{align}
The Riemann tensor is expanded as
\begin{align}
& R^{\mu \rho}_{~~\nu \sigma } = \bar{R}^{\mu \rho}_{~~\nu \sigma } +\frac{2}{l^2} h^{\mu }_{[\nu } \delta^{\rho }_{\sigma ]}
 -2\bar{\nabla}_{[ \nu}\bar{\nabla}^{[\mu}h^{\rho]}_{\sigma ]} +\mathcal{O}(h^2). \label{h-exp_Riemann}
\end{align}
Substitution of (\ref{h-exp_Riemann}) for (\ref{spC_EOM}) unveils the speciality of the present coupling choice: 
The $h$-expansion starts from the $h$-squared term.
Then we calculate $T^{(h)}_{\mu \nu}$ at the lowest order,
\begin{align}
T^{(h)}_{\mu \nu} = 
\tfrac{15 l^2}{4\pi G}~ g_{\mu \nu_5}\delta ^{[ \nu_1}_{[ \nu} \bigl( \tfrac{1}{l^2} h^{\nu_2 }_{\nu_1 } \delta^{\nu_3 }_{\nu_2 }-\bar{\nabla}_{\! \nu_1}\! \bar{\nabla}^{\nu_2}h^{\nu_3}_{\nu_2} \bigr)
 \bigl( \tfrac{1}{l^2} h^{\nu_4 }_{\nu_3 } \delta^{\nu_5 ]}_{\nu_4 ]}-\bar{\nabla}_{\! \nu_3}\! \bar{\nabla}^{\nu_4}h^{\nu_5 ]}_{\nu_4 ]} \bigr).
\end{align}
The Killing charge conservation indicates the existence of the co-exact 2-form $\mathcal{B}$,
\begin{align}
 \bar{\nabla}_{\mu}(T_{(h)}^{\mu \nu} \xi _{\nu}) =0,~~~~~
 T_{(h)}^{\mu \nu} \xi _{\nu} = \bar{\nabla}_{\lambda }\mathcal{B}^{\lambda \mu },
\end{align}
where $\xi$ is a Killing vector field. 
Conserved charges are expressed by $\mathcal{B}$ with the surface integral
\begin{align}
& Q^t(\xi) = \frac{1}{2\pi^2} \int_{\mathcal{M}_4}\! T_{(h)}^{t \nu} \xi _{\nu} \text{d}V =\frac{1}{2\pi^2} \int_{S^3} \mathcal{B}^{r t} \text{d}\Sigma_r.
\end{align}
This 2-form in EGB gravity with the special coupling choice is found out as follows
\begin{align}
 \mathcal{B}^{\mu \nu} = \tfrac{15 l^2}{4\pi G} \bigl( h_{[\nu_1}^{[ \mu }\bar{\nabla}^{\nu}\xi ^{\nu_1} +2\xi^{[\mu }\bar{\nabla}^{\nu}h_{[\nu_1}^{\nu_1 } \bigr)
 \bigl( \tfrac{1}{l^2} h^{\nu_2 }_{\nu_2 } \delta^{\nu_3 ]}_{\nu_3 ]} -\bar{\nabla}_{\! \nu_2 }\! \bar{\nabla}^{\nu_2}h^{\nu_3 ]}_{\nu_3 ]} \bigr).
\end{align}
Adopting this formula seems reliable because this integral actually gives the correct mass value in the static black hole case (\ref{SpCR}).
In our case, it is natural to assign the Killing charge of $\xi = \partial _t$ and $\partial _{\phi} $ as the mass $m$ and the angular momentum $j$ respectively.
These values we evaluated are 
\begin{align}
 m= Q^t(\partial _t) = \frac{l^2 \mu_2 -\mu_1 }{8\pi G},~~~
 j= Q^t(\partial _{\varphi }) = -\frac{ l\mu_1}{16\pi G},
\end{align}
Moreover, the Maxwell charge $q$ can be obtained with the ordinary manner
\begin{align}
q= \frac{1}{2\pi^2} \int_{S^3} \ast F = \frac{Q}{l}.
\end{align}
In higher dimension the straightforward calculation shows that our solutions also have these charges, 
especially acquire the finite value angular momentum.

\section{Discussion}
In this paper, we present new exact solutions of every odd dimension in Einstein Gauss-Bonnet gravity with Maxwell electrodynamics.
Some proficient properties of our solutions are revealed i.e. 
stationarity beyond static, asymptotic AdS at least locally, and most characteristic one, Sasakian structure.
Somewhat intricate relationship is also unveiled between our solutions and already-known solutions 
such as the rotating one only found previously, or the class of static topological black holes.
Evaluating conserved charges with the AD formulation, we confirm that the present spacetime is actually rotating.
This evaluation itself also seems worthful since it clarifies one peculiar property of EGB gravity different from Einstein gravity.

Our manner can be regarded as one of alternatives to Kerr-Schild ansatz 
because the naive Kerr-Schild form cannot catch present solutions.
We are convinced that more other solutions can be caught by the present ansatz or some variations, 
in fact getting some CS case or even-dimensional counterparts already \cite{Future}.
This work also has the contribution to Sasakian geometry itself as a concrete example.
Recently the metric with Sasakian structure has been discussed intensively, 
but concrete solutions in physical theories are so few that only in Einstein gravity.

We consider the necessity for detailed global analysis to improve the present coordinate system.
The reason is that some defects should occur at the Euclidean global viewpoint after the naive Wick rotation of the present coordinate.
This expectation arises from the prior research of $Y^{pq}$, the Euclid Sasaki-Einstein manifold satisfying our ansatz locally.
The global regular condition of its metric has been established well, requiring two local parameters to be discrete values \cite{Gauntlett:2004yd}.
In our present coordinate this condition is certainly broken, 
since it demands at least the radial shift $y \rightarrow y+\chi $ and specific transformations which get into singular if $\chi $ vanishes.
We expect this parameter have some physical meaning.
It is noteworthy that the known metric mentioned as supersymmetric and naked-singular \cite{Cvetic:2005nc} also breaks this condition after the suitable Euclideanization.
The Euclidean consistency seems to be useful for extracting some physics from our solutions.

Several progress is expected at exploring the above analysis, included the purely Euclidean argument.
Because in Euclidean regime the established manner is available for avoiding the singularity and getting the compactness, 
this should be applicable in some scenarios on super string theory as the compact manifold, 
e.g. $\mathcal{M}_5 $ of $\mathrm{AdS_5/CFT_4}$.
In Lorentzian regime, its physical conditions will be improved.
Using the enough large isometry, we want to analyze the instability in the systematic manner \cite{Dotti:2005sq, Murata:2008yx}.
Proceeding the mysterious relationship between Sasakian manifolds and Kerr-NUT-AdS spacetimes \cite{Hashimoto:2004ks, Houri:2008th}, 
we will contribute to the emergence of both fields.

\acknowledgments
We would like to thank Hideaki Aoyama, Tatsuo Kobayashi, Jiro Soda, Yukinori Yasui, Katsuyuki Sugiyama and Tsuyoshi Houri 
for useful comments and illuminating discussions.
This work is supported by the Grant-in-Aid for the Global COE Program 
"The Next Generation of Physics, Spun from Universality and Emergence" 
from the Ministry of Education, Culture, Sports, Science and Technology (MEXT) of Japan.

\end{document}